# THE FRENCH NATIONAL 3D DATA REPOSITORY FOR HUMANITIES: FEATURES, FEEDBACK AND OPEN QUESTIONS

Sarah Tournon [1], Vincent Baillet [1], Mehdi Chayani [1], Bruno Dutailly [1,2],
Xavier Granier [3], Valentin Grimaud [4,5]

1 Archeovision (UMS CNRS, Université de Bordeaux Montaigne, Université de Bordeaux)
2 PACEA (UMR CNRS, Université de Bordeaux, Ministère de la Culture)
3 LP2N (UMR CNRS, Institut d'Optique Graduate School, Université de Bordeaux)
4 LARA – CreAAH (UMR CNRS, Universités du Mans, de Nantes, Rennes1, Rennes2, Ministère de la Culture)
5 UFR HHAA - Université de Nantes - UFR Histoire, Histoire de l'Art et Archéologie

*Abstract*

We introduce the French National 3D Data Repository for Humanities designed for the conservation and the publication of 3D research data in the field of Humanities and Social Sciences. We present the choices made for the data organization, metadata, standards and infrastructure towards a FAIR service. With 437 references at the time of the writing, we have feedback on some challenges to develop such a service and to make it widely used. This leads to open questions and future developments.

## 1. Introduction and Context

Research in the field of Humanities and Social Sciences (HSS) are integrating the use of 3D data in their scientific practice.

With the democratization of 3D technologies, the increasing number of 3D data raises several questions (Münster *et al.* 2016) around a major one: how to ensure data transmission for future research? The resulting questions concern the archiving, conservation or publication processes, and thus the data format, the standard metadata and the infrastructure.

Archiving digital data is an issue for research and Cultural Heritage institutes (Alliez *et al.* 2017) and the number of initiatives are growing in the field of cultural and, more broadly, archaeological heritage archive management. Widely mentioned in the UNESCO Charter for the preservation of digital heritage (UNESCO 2003), this challenge is more specifically raised for the digitisation of Cultural Heritage in the London Charter (Denard 2009). However, archiving will be addressed only in 2014 by the European Archaeological Council (Perring *et al.* 2014) and updated in 2021 (Onizsczuk *et al.* 2021). If we are looking for dedicated and existing solutions for 3D data in archaeology, we must rely on national initiatives such as ADS in United Kingdom (ADS 2014). In France, the Archeogrid project, publicly launched in 2005 (Vergnieux 2005), provided the environment to develop the French National 3D Data Repository for Humanities (CND3D 2018).

To achieve such a development, a consortium "3D for Humanities" (CST3D 2021) has been created in 2014 as a scientific network of 3D models producers and users in the field of archeology and Cultural Heritage. It was created with the support of French national digital infrastructure for Human and Social Sciences (Huma-Num 2013). The participants have defined guidelines for the conservation and publication of 3D data, with all related documents and well-defined metadata scheme (Dutailly et al. 2019), within the goal of a FAIR service (Wilkinson *et al.* 2016).

*Figure 1: A deposit in the National 3D Data Repository for Humanities with one virtual object*
*(DOI: 10.34969/CND3D/257350.d.2015)*

Nowadays, where a data management plan is essential to submit a project, the Conservatory's vocation is still to provide a solution to the strategic needs of the long-term preservation of research data as well as their diffusion.

In this paper, we present an overview of its features together with some first feedback based on now 3 years' experience.

## 2. Features

As illustrated in Figure1, the National 3D Data Repository is organized around the two notions of

1. Virtual object: a 3D digital instance of a coherent entity, from a Cultural Heritage point of view: that has been digitised and/or restituted (Granier et al. 2021). Each virtual object includes not only the final model, but also to all the documents (from digital scans, images, to texts …) that have been used.

2. Deposit: a collection of coherent virtual objects.

All the documents, the virtual object, and the deposit are associated with corresponding metadata (Dutailly et al. 2019, see also Figure 3).

### 2.a Findable and Accessible

Publishing data means making them available to the scientific community. It is therefore necessary to ensure that documentation and metadata complies with interoperable standards. We use Dublin Core (DCMI 2012) with an ongoing alignment with the CIDOC-CRM (CIDOC CRM SIG 2021) to integrate the European ARIADNE portal (Niccolucci and Richards 2019). While keeping the possibility of defining its own referential, the National 3D Data Repository is aligned with PeriodO (Golden and Shaw 2016), Geonames (Bond and Bond 2019) and the PACTOLS thesauri (Nouvel 2019) from the OpenTheso platform (Rousset 2005).

The metadata can be filled on-line, or off-line with a specific open-source software named aLTAG3D "a Long-Term Archive Generator for 3D" (Dutailly and Gaillardo 2021). This software allows us to construct the package, object-by-object by attaching the sources to it and by filling in, at all levels the requested metadata. It was designed to simplify the tedious task of metadata filling for such complex data.

### 2.b Interoperable and Reusability

A DOI is assigned at deposit and virtual object levels. This DOI allows publications in the open archive HAL (Baruch 2007, CCSD 2021) to point the referenced 3D data. Reciprocally, the related publications in HAL are referenced by their unique identifier.

The ultimate goal of FAIR is to optimize the reuse of data. Standard 3D formats PLY (Turk 1994) and DAE (Barnes and Levy Finch 2008) and non-standard formats are allowed. By using recognized standards formats, data and metadata can be pushed to a national long term archiving facility (CINES 2021). Using non-supported standards or non-standard formats prevents any full archiving possibility, but we have chosen to offer such a service to spread the use of such a facility.

### 2.c Storage

We use a national networked distributed, long-term, and secured storage facilities provided by Huma-Num. To preserve the accessing right on sensitive data, authentication and right management are included.

## 3. Feedback

Creating new deposits and references to virtual objects requires the same amount of work than adding a classical publication on an open archive, despite a few more dedicated metadata in order to be referenced for use in Human and Social Sciences.

However, the time spent to actually deposit each object can be quite large, according to the data complexity since it contains not only 3D files, but also all the related documents. The quantity, as well as the variety of metadata is increasing with the different files, which can considerably increase the time required for a data record. Despite that aLTAG3D simplifies the process and is a real asset thanks to its easy handling and architecture, we have to develop more tools which ease data deposit, including plug-ins to automatically retrieve already existing metadata.

Another major problem concerns the interoperability of 3D data. If we archive our 3D models in FBX, GLTF or OBJ formats, we will encounter a real difficulty when we want to run these data within the 3D software commonly used in the humanities and social sciences. Indeed, 3D models are linked to textual data, plans, architectural surveys, textures, etc. The bridges established during the construction of a 3D model tend to disappear partially or totally, when the scenes are opened again with the formats indicated above. This phenomenon is even more visible when opening these data in a software that is different from the original that allowed the constitution of the 3D data. It would be useful to develop an in-house script to re-establish the links between the 3D model and these peripheral data. In addition, there are already paid scripts that facilitate the opening of 3D files between different software. In our opinion, it is important to avoid this thorny problem of opening and accounting for 3D data. Because if we can ensure the archiving of our 3D data, it is also important that we can consult them easily without losing useful data to understand the 3D.

**4. Open Questions**

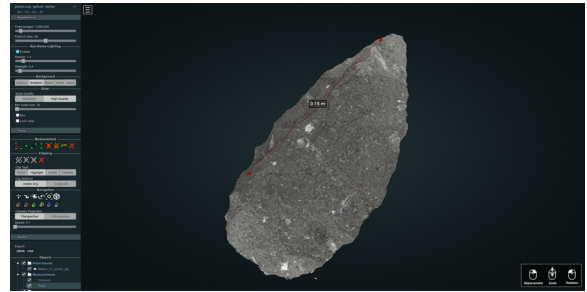

*Figure 2: Online 3D visualisation of a point cloud using Potree viewer ([DOI:10.34969/CND3D/500986.d.2021](DOI:10.34969/CND3D/500986.d.2021)).*

The main challenge for the wide acceptance of 3D model archiving remains to ease the filling process of metadata for such complex data. Several avenues can be explored to address this constraint.

Firstly, the number of metadata to be filled in can be a source of demotivation. It is necessary to find the right balance between the just amount of pertinent data to create, with a reduced but well documented amount of information. This point illustrates even more that any digitisation project needs to integrate a data management plan.

Secondly, another perspective would be not to limit the creation of a repository to archiving only, but to multiply the possible uses. For example, the creation of an archive package can be the source of an interactive 3D visualisation through a web browser. Among existing technologies (Scopigno et al. 2017, Vurpillot et al. 2018, Fung et al. 2021), we are experimenting 3DHOP (Potenziani et al. 2015) for meshes and Potree (Schütz 2016) for point clouds (see Figure 2).

One of the main challenges remains the creation of a perennial services while minimizing the required to be increased storage cost that has an economic and ecological impact that we cannot neglect for such data sizes. Indeed, the tool is currently designed to bring all data together in the same repository. However, some resources may already be deposited on another service with a medium or long-term conservation policy (Nakala for example). Data redundancy is therefore not an optimal way of using the different storage spaces, and this represents a significant cost for the hosts, and our planet. One approach would therefore be to integrate data into the repository as external references – provided that the continuity of the link is guaranteed. This comes together with the always increasing possible interconnections with other repositories using newly defined or new metadata standards and will increase the visibility of the indexed objects.

In this way, the documented data can be harvested by search engines. The 3D models are therefore no longer stored and forgotten in a storage space but can be called upon at any time for research, teaching or even mediation purposes towards an increasingly demanding public.

Finally, we must also admit that the metadata scheme in its current state struggles to cover all the uses of 3D models. Evolutions must therefore be made to adapt to increasingly varied technologies and scientific objectives. This covers, for example, the convergence between 3D models and GIS, the integration in immersive environments (virtual, augmented, or mixed reality), and all types of data associated with these applications.

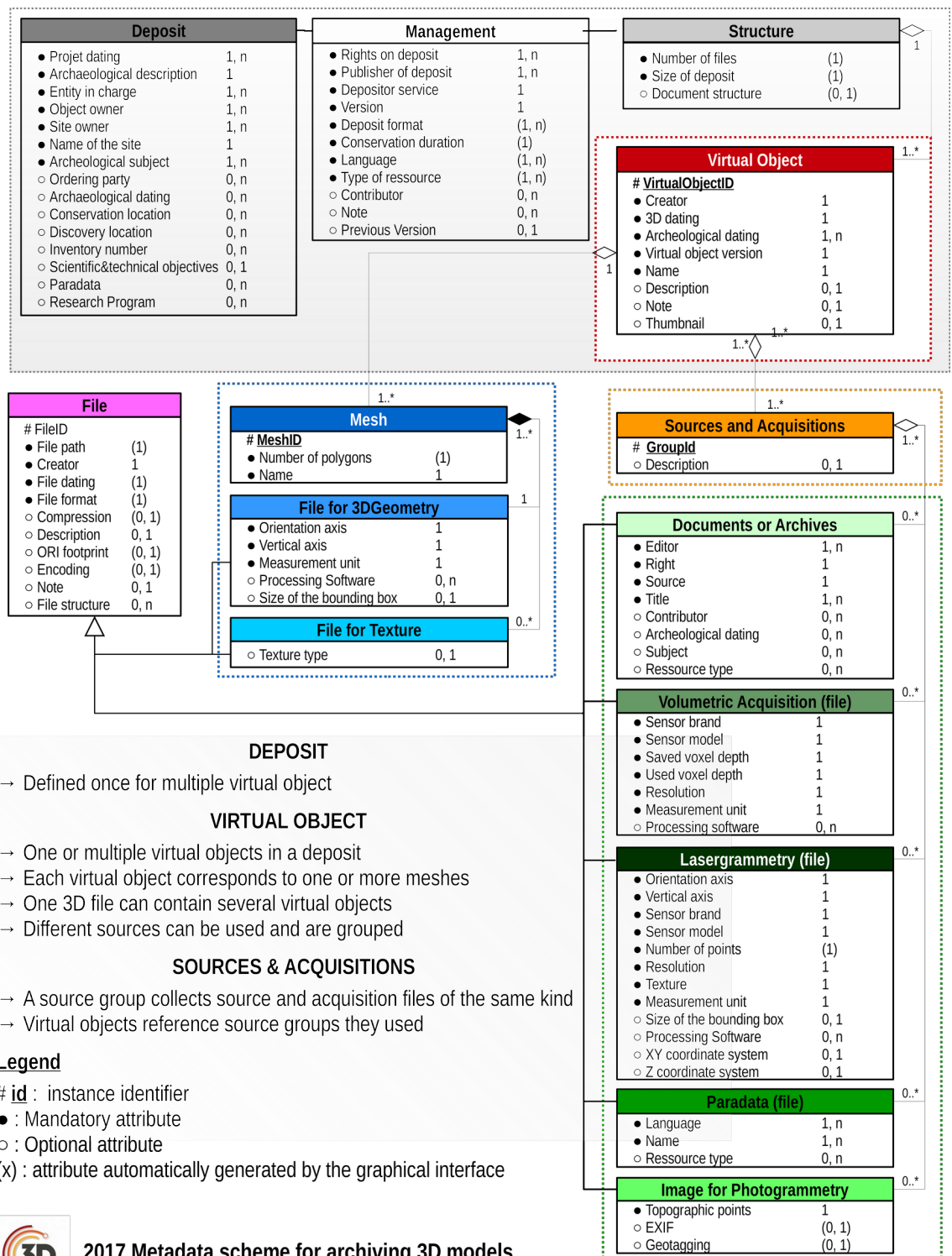

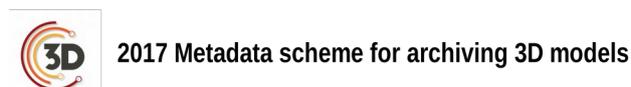

*Figure 3: Translated metadata structure proposed by the consortium "3D for Humanities".*